\documentclass[12pt]{article}
\usepackage{mathrsfs}
\usepackage{amsmath,amssymb}
\usepackage{indentfirst}
\usepackage{amsfonts}
\usepackage{bm}
\usepackage{graphicx}
\usepackage{natbib}
\usepackage{authblk}

\newtheorem{cor}{Corollary}[section]

\newtheorem{thm}{Theorem}[section]

\newtheorem{asp}{Assumption}[section]

\title{Non-parametric shrinkage mean estimation for quadratic loss functions with unknown covariance matrices}
\author[1,2]{Cheng Wang}
\author[2]{Tiejun Tong}
\author[3]{Longbing Cao}
\author[1]{Baiqi Miao}
\affil[1]{Department of Statistics and Finance, University of Science and Technology of China, Hefei, Anhui 230026, China}
\affil[2]{Department of Mathematics, Hong Kong Baptist University, Hong Kong}
\affil[3]{Advanced Analytics Institute, University of Technology, Sydney, NSW 2007, Australia}

\date{\today}
\begin{document}
\maketitle

\begin{abstract}
In this paper, a shrinkage estimator for the population mean is proposed under known quadratic loss functions with unknown covariance matrices.
The new estimator is non-parametric in the sense that it does not assume a specific parametric distribution for the data
and it does not require the prior information on the population covariance matrix.
Analytical results on the improvement of the proposed shrinkage estimator are provided and some corresponding asymptotic properties are also derived.
Finally, we demonstrate the practical improvement of the proposed method over existing methods through extensive simulation studies and real data analysis.

\textit{Keywords:} High-dimensional data; Shrinkage estimator; Large $p$ small $n$; $U$-statistic.
\end{abstract}

\section{Introduction}

\noindent
High-throughput molecular technologies that enable researchers to collect and monitor information at the genome level have revolutionized the field of biology in the past fifteen years. These data offer an unprecedented amount and diverse types of data that reveal different aspects of the biological processes.
One such example is microarray data, where the expression levels of thousands of genes are measured simultaneously from each sample.
These data have motivated the development of reliable biomarkers for disease subtypes classification and diagnosis,
and for the identification of novel targets for drug treatment.
Due to the cost and other experimental difficulties such as the availabilities of biological materials,
it is common that high-throughput data are collected only in a limited number of samples.
They are often referred to as high-dimension, low-sample-size data, or ``large $p$ small $n$" data where $p$ is the number of genes or dimensions and $n$ is the sample size.
High-dimensional data pose many challenges to the traditional statistical and computational methods.
Specifically, due to the small size $n$, there are more uncertainties associated with standard estimators of parameters such as the mean and variance estimations.
As a consequence, statistical analyses based on such parameter estimation are usually unreliable.

In this work, our interest is conducing a more accurate estimator for the population mean $\mu$ under the ``large $p$ small $n$" setting (\citealp{HwangLiu2010,tong2012improved}).
An accurate estimate of $\mu$ is desired in many areas of statistical analysis, e.g., in linear discriminant analysis (\citealp{anderson2003}),
diagonal linear discriminant analysis \citep{dudoit2002comparison}, Markowitz mean-variance analysis \citep{markowitz1952portfolio,el2010high} and so on. Under the assumption that $\mu$ is sparse, \cite{shao2011sparse} proposed a consistent estimator for $\mu$ under some regular conditions. However, in many real problems, there is often little prior information on $\mu$ and it may not necessarily have a sparse structure.
In such situations, the shrinkage estimation of $\mu$ can be applied.
Shrinkage estimation starts with the amazing result of \cite{james1961estimation} that
the commonly used sample mean of a normal distribution is inadmissible and can be improved by shrinkage estimators.
We refer to them as James-Stein type estimators.
Since then, there is a large body of literature in shrinkage estimation including \cite{baranchik1970family}, \cite{efron1973stein}, \cite{lin1973generalized}, \cite{berger1977minimax}, \cite{gleser1986minimax}, \cite{fourdrinier2003robust}, \cite{chetelat2012improved} and etc.
In the literature, most existing methods either assumed that the covariance matrix $\Sigma_p$ is known
or assumed that there exists an estimator of $\Sigma_p$ that is invertible.
As a common practice, if the sample covariance matrix $S_n$ is used to estimate $\Sigma_p$,
the sample size is required to be larger than the dimension, i.e. $n>p$, to avoid the singularity.
Note that, however, for high-dimensional data it is common that $p$ is much larger than $n$.
To overcome the singularity problem for high-dimensional data, \cite{tong2012improved} proposed a new shrinkage estimator for $\mu$ by assuming that $\Sigma_p$ has a diagonal structure which may not be realistic. Therefore, the traditional shrinkage methods can not be applied to analyze high-dimensional data directly.

Inspired by \cite{ledoit2004well}, in this paper we consider the shrinkage estimation for $\mu$ under quadratic loss functions
with unknown non-diagonal covariance matrices.
The new estimator is non-parametric in the sense that it does not assume a specific parametric distribution for the data
and it does not require the prior information on covariance matrix $\Sigma_p$.
We will demonstrate by both theoretical and empirical studies that the proposed estimator has good properties for a wide range of settings.
We will also show that the proposed method is better than the sample mean and the existing shrinkage methods even under a diagonal covariance matrix assumption.

The rest of the paper is organized as follows. Section 2 introduces the theoretical optimal shrinkage estimation under quadratic risks.
Section 3 develops a data-driven shrinkage estimator and derives the asymptotic properties of the proposed estimator.
We then conduct simulation studies on simulated data in Section 4 and using real data in Section 5 to evaluate the proposed optimal shrinkage estimator
and compare it with existing shrinkage methods.
Finally, we conclude the paper in Section 6 and provide the technical results in the Appendix.

\vskip 12pt
\section{Methodology}

\noindent
Let $X_1,\cdots,X_n$ be independent and identically distributed (i.i.d.) observations satisfying the multivariate model
\begin{eqnarray} \label{data}
X_i=\Sigma_p^{1/2} \epsilon_i+\mu, ~~~~~i=1,\cdots,n,
\end{eqnarray}
where $\mu$ is a $p$-dimensional vector, $\Sigma_p$ is a positive definite matrix and the random errors in
$(\epsilon_{ij})_{p \times n}=(\epsilon_1,\cdots,\epsilon_n)$ are i.i.d. with zero mean, unit variance and finite fourth moment.
Note that model (\ref{data}) has been widely used in the literature such as \cite{bai1996effect} and \cite{chen2010tests}.
In this paper, we do not assume that the data follow a multivariate normal distribution with mean $\mu$ and covariance matrix $\Sigma_p$.
Given model (\ref{data}), we consider to estimate $\mu$ under the following quadratic loss function \citep{berger1976admissible, berger1977minimax, gleser1986minimax},
\begin{eqnarray} \label{loss}
L_Q(\delta)=n (\delta-\mu)'Q(\delta-\mu)/{\rm tr}(Q\Sigma_p),
\end{eqnarray}
where $\delta=\delta(X_1,\cdots,X_n)$ is the estimator of $\mu$, $Q$ is a known positive definite matrix, and $\rm{tr}(A)$ stands for the trace of matrix $A$.
Note that for the standard sample mean $\bar{X}=(1/n)\sum_{k=1}^n X_k$, the risk function is $E[L_Q(\bar{X})]= 1$.

In the special case when $X_1,\cdots,X_n$ are multivariate normal distributed, \citet{james1961estimation} showed that
\begin{eqnarray}
\delta_{JS}=(1-\frac{p-2}{n \bar{X}'\bar{X}})\bar{X}
\end{eqnarray}
dominates $\bar{X}$ for any $p>2$ under the assumption that $\Sigma_p=Q=I_p$.
This result was then extended by \cite{baranchik1970family} to $\Sigma_p=\sigma^2I_p$ with $\sigma^2$ unknown,
and by \citet{efron1973stein} to a Bayesian estimator.
For a general unknown $\Sigma_p$, the James-Stein estimator has the form (\citealp{lin1973generalized, berger1976admissible, berger1977minimax, gleser1986minimax, fourdrinier2003robust})
\begin{eqnarray}
\delta_{JS}=(I-\frac{r(Q,S_n^{-1},\bar{X})}{\bar{X}'S_n^{-1}\bar{X}})\bar{X},
\end{eqnarray}
where $r(Q,S_n^{-1},\bar{X})$ is a measurable function of $Q$, $S_n^{-1}$ and $\bar{X}$ with
$0 \leq r(Q,S_n^{-1},\bar{X}) \leq 2(G-2)/(n-G+2)$ and $S_n=\sum_{k=1}^n (X_k-\bar{X})(X_k-\bar{X})'/(n-1)$ being the sample covariance matrix.
To guarantee $S_n$ is invertible, $n>p$ is necessary which means the method is not applicable for the ``large $p$ small $n$" data.

To overcome the singularity problem, \cite{tong2012improved} considered a special situation where $\Sigma_p$ is diagonal.
Specifically, under the loss function with $Q=\Sigma_p^{-1}$ they constructed a hierarchical Bayesian model and then proposed the following shrinkage estimator,
\begin{eqnarray}
\delta_{T}=(1-\frac{(p-2)(n-1)}{n(n-3) \bar{X}'D_{n}^{-1} \bar{X}})\bar{X}
\end{eqnarray}
where $D_{n}={\rm diag}(S_n)$.
Other related works for a diagonal $\Sigma_p$ and a diagonal $Q$ assumptions include \cite{berger1976combining} and \cite{shinozaki1980estimation}.
Whereas for an arbitrary $Q$ with non-diagonal $\Sigma_p$, it remains a challenging yet unanswered question under the ``large $p$ small $n$" setting.
To address this question, we consider to estimate $\mu$ by a linear combination of $\bar{X}$ and $e=(1,\cdots,1)'$,
\begin{eqnarray*}
\delta=\alpha \bar{X}+\beta e.
\end{eqnarray*}
The following theorem derives the optimal shrinkage coefficients for model (\ref{data}) under the quadratic loss (\ref{loss}) with an arbitrary known $Q$.

\vskip 12pt
\begin{thm} \label{thm1}
Consider the optimization problem,
\begin{eqnarray}
\min_{\alpha,\beta} E (\delta-\mu)'Q(\delta-\mu)~~~{\rm s.t.}~~~\delta=\alpha \bar{X}+\beta e,
\end{eqnarray}
where the coefficients $\alpha$ and $\beta$ are non-random.
The optimal shrinkage estimator is given as $\mu^*=\alpha^* \bar{X}+\beta^* e$ where
\begin{eqnarray*}
&& \alpha^*=\frac{\mu' Q \mu-\frac{(e' Q \mu)^2}{e' Q e}}{\mu'Q \mu +\frac{1}{n}
{\rm tr}(Q \Sigma_p)-\frac{(e' Q \mu)^2}{e'Qe}},\\
&&\beta^* =\frac{\frac{1}{n} {\rm tr}(Q \Sigma_p)}{\mu'Q \mu +\frac{1}{n} {\rm tr}(Q
\Sigma_p)-\frac{(e' Q \mu)^2}{e'Qe}} \frac{e' Q \mu}{e'Q e},
\end{eqnarray*}
and the corresponding risk of $\mu^*$ is
\begin{eqnarray}
E (L_Q(\mu^*))=\frac{(\mu-\frac{e' Q \mu}{e' Q e}e)' Q
(\mu-\frac{e' Q \mu}{e' Q e}e)}{(\mu-\frac{e' Q \mu}{e' Q e}e)' Q (\mu-\frac{e'
Q \mu}{e' Q e}e) +\frac{1}{n} {\rm tr}(Q \Sigma_p)}.
\end{eqnarray}
\end{thm}

\vskip 12pt
Note that the proposed shrinkage estimator can accommodate any shift of the grand mean, including the shift from $\mu$ to $\mu+c e$ where $c$ is a constant.
This is a similar idea as that in \cite{stein1962discussion} where the author shrunk the observations to grand mean rather than to the origin.
Also in \cite{tong2012improved}, the authors applied their shrinkage method to the grand mean and so the final estimator was a linear combination of two different components.
By Theorem \ref{thm1}, however, we point out that the method in \cite{stein1962discussion} is not applicable for arbitrary $Q$.
For this point, we will explain in the simulation study an example where the grand mean is zero but $e' Q \mu \neq 0$.

\vskip 12pt
\section{Data-driven shrinkage estimators for population means}

\noindent
Note that the shrinkage coefficients $\alpha^*$ and $\beta^*$ are unknown and need to be estimated in practice.
In this section, we propose to estimate them by $U$-statistics, motivated from \cite{chen2010tests}, \cite{cai2012optimal} and \cite{li2012two}.
Specifically, we estimate $\alpha^*$ and $\beta^*$ by
\begin{eqnarray*}
\hat{\alpha}^*=\frac{Y_{1,n}-Y_{3,n}}{Y_{1,n}+Y_{2,n}-Y_{3,n}} ~~~{\rm and}~~~ \hat{\beta}^*=\frac{Y_{2,n}}{Y_{1,n}+Y_{2,n}-Y_{3,n}} Y_{4,n}
\end{eqnarray*}
where
\begin{align*}
Y_{1,n} &= \frac{1}{p(n-1)} \sum_{i \neq j} X'_i Q X_j, \\
Y_{2,n} &= \frac{1}{n p} (\sum_{k=1}^n X'_k Q X_k-\frac{1}{n-1} \sum_{i \neq j} X'_i Q X_j ),\\
Y_{3,n} &= \frac{1}{ p (n-1)e' Q e } \sum_{i\neq j} e' Q X_i X'_j Q e, \\
Y_{4,n} &= \frac{1}{n e' Q e } \sum_{k=1}^n e' Q X_k.
\end{align*}
The resulting estimator of $\mu$ is then $\hat{\mu}^*=\hat{\alpha}^* \bar{X}+\hat{\beta}^* e$.
To derive the asymptotic properties of the proposed estimator, we need the following regularity condition.

\begin{asp}\label{con1}
There is a constant $c_0$ (not
depending on $p$ or $n$) such that
\begin{eqnarray*}
c_0^{-1} \leq all~eigenvalues~of~\Sigma_p~and~Q \leq c_0
\end{eqnarray*}
\end{asp}

Under Assumption \ref{con1}, we have ${\rm tr}(\Sigma_p Q)/p=O(1)$. In this work, $o(1)$ denotes
a sequence of random variables that converges to zero and $O(1)$ is short for a sequence that is bounded.
Similarly, $o_p(1)$ and $O_p(1)$ are notations in probability. For more details, one may refer to \cite{van2000asymptotic}.
Let
\begin{eqnarray*}
&&\pi_1=E (\bar{X}-\mu)'Q(\bar{X}-\mu)=\frac{1}{n}{\rm tr}(Q\Sigma_p),\\
&& \pi_2=(\mu-\frac{e'Q \mu}{e'Q e}e)'Q(\mu-\frac{e'Q \mu}{e'Q e}e).
\end{eqnarray*}
The following theorems establish the rates of convergence for the proposed estimators and for the loss function.

\vskip 12pt
\begin{thm} \label{lem1}
Under Assumption \ref{con1},
\begin{eqnarray*}
	&&Y_{1,n}=\frac{n}{p} \mu' Q \mu+O_p(\frac{1}{\sqrt{p}}+\frac{\sqrt{n \mu'\mu}}{p}),~~~~~Y_{2,n}=\frac{1}{p}
	{\rm tr}(\Sigma_p Q)+O_p(\frac{1}{\sqrt{np}}),\\
	&&Y_{3,n}= \frac{n (\mu' Q e)^2}{p e' Q e}+O_p(\frac{1}{p}+\frac{\sqrt{n \mu'\mu}}{p}),~~~~~Y_{4,n}=\frac{e'Q
		u}{e'Q e}+O_p(\frac{1}{\sqrt{np}}).
\end{eqnarray*}
Further, we have
\begin{eqnarray*}
\hat{\alpha}^*=\alpha^*+O_p(\frac{1}{\sqrt{p}}+\frac{\sqrt{n \mu'\mu}}{p}), ~~~{\rm and}~~~
\hat{\beta}^*=\beta^*+O_p(\frac{1}{\sqrt{np}}+\frac{\sqrt{n} \mu' \mu}{p \sqrt{p}}).
\end{eqnarray*}
Therefore, $\hat{\alpha}^*-\alpha^* \buildrel p \over \rightarrow 0$ and $\hat{\beta}^*-\beta^* \buildrel p \over \rightarrow 0$
as $p \to \infty$ and $ (n p^{-2}+n^{1/2} p^{-3/2}) \mu'\mu \to 0$, where $\buildrel p \over \rightarrow$ denotes convergence in probability.
\end{thm}

\vskip 12pt
\begin{thm} \label{thm3}
Under Assumption \ref{con1}, under the ``large $p$ small $n$" setting the loss function of the shrinkage estimator $\hat{\mu}^*$ is
\begin{eqnarray}
L_Q(\hat{\mu}^*) =\frac{\pi_2}{\pi_1+\pi_2}+O_p(\frac{1}{\sqrt{p}}+\frac{\sqrt{n \mu' \mu}}{p}).
\end{eqnarray}
\end{thm}

By Theorem \ref{thm3}, we note that $\hat{\mu}^*$ behaves at least as well as $\bar{X}$ when $p$ is large and $n p^{-2} \mu' \mu \to 0$.
The explicit improvement of $\hat{\mu}^*$ over $\bar{X}$ depends on $\pi_1$ and $\pi_2$.
As in \cite{ledoit2004well}, we define the percentage relative improvement in average loss (PRIAL) over the sample mean as
\begin{eqnarray}
\rm{PRIAL}=\frac{E
(\bar{X}-\mu)'Q(\bar{X}-\mu)-(\hat{\mu}^{*}-\mu)'Q(\hat{\mu}^{*}-\mu)}{E
(\bar{X}-\mu)'Q(\bar{X}-\mu)}.
\end{eqnarray}
We then have the following corollary.

\vskip 12pt
\begin{cor} \label{cor}
Let $s_n=\frac{n}{p}(\mu-\frac{e'Q \mu}{e'Q e}e)'Q(\mu-\frac{e'Q \mu}{e'Q e}e)$. As $p \to \infty$ we have
\begin{itemize}
\item[(I)] If $s_n \to 0$,  ${\rm PRIAL} \buildrel p \over \rightarrow 1$;
\item[(II)] If $s_n \to C_0$,  ${\rm PRIAL} \buildrel p \over \rightarrow C_1 \in
(0,1)$;
\item[(III)] If $s_n \to \infty $,  ${\rm PRIAL} \buildrel p \over \rightarrow 0$.
\end{itemize}
\end{cor}
Therefore, the shrinkage estimator $\hat{\mu}^*$ always performs better than $\bar{X}$ under the loss function (\ref{loss}) when $s_n$ is finite.
In the extreme case when $s_n \to \infty$, $\hat{\mu}^*$ behaves similarly as $\bar{X}$.

\vskip 12pt
\section{Simulation studies} \label{part4}
\noindent
In this section, we conduct simulation studies to evaluate the performance of the proposed shrinkage estimator $\hat{\mu}^*$
and compare it with the following four estimators: the James-Stein estimator  $\delta_{JS}$ in \cite{baranchik1970family},
the Berger-Bock estimator $\delta_{BB}$ in \cite{berger1976combining}, the Tong et al. estimator $\delta_T$ in \cite{tong2012improved} and the CW estimator $\delta_{CW}$ proposed by \cite{chetelat2012improved}.

Our first simulation will be conducted under the classical situation \citep{james1961estimation} where $Q=\Sigma_p=I_p$. In particular, we considered two
scenarios with respect to the innovation random vector $\epsilon_{ij}$ in model (\ref{data}):
\begin{enumerate}
\item[(I)] $\epsilon_{ij}$ are i.i.d from standard normal distribution that the data are distributed from Gaussian data with mean $\mu$ and an identity covariance matrix;

\item[(II)] $ \sqrt{v/(v-2)} \times \epsilon_{ij}$ are i.i.d from $t$-distribution with $v$ degrees of freedom where the constant $ \sqrt{v/(v-2)}$ guarantees that $\epsilon_{ij}$ has zero mean and unit variance.
\end{enumerate}
In both scenarios, we set $\mu \sim N_p(0, I_p)$ and all the simulation results are based on 10,000 simulations.

\vskip 12pt
\begin{table}
\centering
\begin{tabular}{ccccccc}
\hline
$n$&Sample Mean&James-Stein&Berger-Bock&Tong et al.&CW&Proposed\\
\hline
\multicolumn{7}{c}{Normal random vectors}\\
10&1.0012&0.9088&0.9348&0.9094&0.9945&0.9090\\
20&0.9997&0.9508&0.9569&0.9509&0.9913&0.9508\\
50&0.9988&0.9790&0.9799&0.9790&0.9899&0.9790\\
\multicolumn{7}{c}{$t$-distribution with $v=5$}\\
10&1.0011&0.9079&0.9604&0.9097&0.9902&0.9080\\
20&1.0000&0.9511&0.9686&0.9515&0.9893&0.9511\\
50&1.0014&0.9814&0.9854&0.9815&0.9917&0.9815\\
\multicolumn{7}{c}{$t$-distribution with $v=10$}\\
10&0.9998&0.9075&0.9458&0.9080&0.9924&0.9076\\
20&1.0006&0.9519&0.9626&0.9520&0.9919&0.9520\\
50&0.9979&0.9776&0.9796&0.9776&0.9880&0.9776\\
\hline
\end{tabular}
\caption{Empirical risks of the estimators under classical situation where $Q=\Sigma_p=I_p$ and $p=100$.}
\label{tab0}
\end{table}
Table \ref{tab0} reports the empirical risks of the proposed estimator and its competitors for both scenarios
of distributions. We observe from Table \ref{tab0} that overall the James-Stein estimator has the best performances since it utilized the fact that the population covariance matrix is an identity matrix and the proposed estimator has similar average losses as ones of the James-Stein estimator. All the estimators are applicable for variables with normal distributions and $t$-distributions. Moreover, the empirical risks for variables with $t$-distributions follow very similar patterns to those of Gaussian data. This is understandable as for data structure (\ref{data}), there is no too much differences between Gaussian data and random data with other distributions such as $t$-distributions. For this reason, in the following part, we only report the simulation results for Gaussian data.

Note that the existing competitors for comparison, $\delta_{JS}$, $\delta_B$ and $\delta_T$, only work on a diagonal covariance matrix under the ``large $p$ small $n$" setting.
Whereas for the proposed estimator $\hat{\mu}^*$, it works for both diagonal and non-diagonal covariance matrices.
Thus in the following part, for a meaningful comparison, we will consider the quadratic loss function in \cite{tong2012improved}.
Specifically, by letting $Q$ be diagonal and let $Q^{-1}=\rm{diag}(\Sigma_p)$, we have the following loss function,
\begin{eqnarray} \label{risk1}
L(\delta)=\frac{n}{p}(\delta-\mu)'[\rm{diag}(\Sigma_p)]^{-1} (\delta-\mu),
\end{eqnarray}
where the constant $n/p$ is applied to guarantee that $E[L(\bar{X})]=1$. In applications, $Q$ will be estimated from the diagonal elements of the sample covariance matrix.

We will simulate $X_1,\cdots, X_n$ independently from a $p$-dimensional multivariate normal distribution with mean $\mu$ and covariance matrix $\Sigma_p$.
For $\mu$, we consider two options:
\begin{enumerate}
\item[(a)] $\mu_1=(\mu_{11},\cdots,\mu_{1p})'$ where $\mu_{11},\cdots,\mu_{1p}$ are i.i.d from $N(0,\tau^2)$;
\item[(b)] $\mu_2=(\mu_{21},\cdots,\mu_{2p})'$ where $\mu_{2k}=\tau$ for $k \leq p/2$ and $\mu_{2k}=-\tau$ for $k>p/2$.
\end{enumerate}
In both options, we consider $\tau=0.5$ and $1$ to represent different levels of mean heterogeneity. For $\Sigma_p$, we consider three covariance matrices:
\begin{enumerate}
\item[(1)] $\Sigma_1$ is diagonal with 20\% of population eigenvalues being equal to 1, 40\% begin equal to 3 and 40\% being equal to 10;
\item[(2)] $\Sigma_2=\Sigma_1^{1/2} \Sigma_0 \Sigma_1^{1/2}$ where $\Sigma_0=(\sigma_{ij})_{p \times p}$ and $\sigma_{ij}=\rho^{|i-j|}$ for $1 \leq i,j \leq p$;
\item[(3)] $\Sigma_3=\Sigma_1^{1/2} \Sigma_{00} \Sigma_1^{1/2}$ where $\Sigma_{00}=(\sigma_{ij})_{p \times p}$ and $\sigma_{ij}=1$ for $i=j$, $\sigma_{ij}=\rho$ for $i \neq j$.
\end{enumerate}
We let $\rho$ range from 0.1 to 0.9 for $\Sigma_2$ and from 0.1 to 0.5 for $\Sigma_3$ to represent four different levels of dependent structure.

The second simulation study is to evaluate the performance of $\hat{\mu}^*$ with existing methods when $\Sigma_p=\Sigma_1$, i.e., when the covariance matrix is diagonal.
Let $p=100$ throughout the simulations. We consider $n=10$, $25$, $50$ and $100$, to represent different levels of sample sizes.
Table \ref{tab1} reports the empirical risks of the estimators under various settings.
First of all, we observe that all shrinkage methods have a smaller risk than the sample mean $\bar{X}$.
This shows that for high-dimensional data, the shrinkage estimators do improve the standard estimation.
Among the shrinkage estimators, $\delta_T$ and $\hat{\mu}^*$ are among the best in most settings.
The James-Stein estimator $\delta_{JS}$ is not very compatible because it is restricted to a common variance assumption,
and $\delta_{B}$ is only applicable for large sample sizes.
Finally, for $\delta_T$ and $\hat{\mu}^*$, we note that they perform similarly when $\mu=\mu_1$, and $\hat{\mu}^*$ is better by a large margin than $\delta_T$ when $\mu=\mu_2$.
In addition, when the mean heterogeneity increases from $\tau=0.5$ to $\tau=1$,
the improvement of $\hat{\mu}^*$ over $\bar{X}$ decreases which is consistent with Corollary \ref{cor}.
We also observe that the improvements of the shrinkage estimators over the sample mean become smaller when $n$ becomes larger.
This is meaningful since for the large sample size scenario, the mean estimation itself is good enough
and it is no longer necessary to borrow information from others to improve the estimation.

\vskip 12pt
\begin{table}
\centering
\begin{tabular}{cccccccc}
\hline
$\mu$&$\tau$&$n$&Sample Mean&James-Stein&Berger-Bock&Tong et al.&Proposed\\
\hline
$\mu_1$&0.5&10&1.0064&0.5160&0.6175&\textbf{0.4763}&0.4867\\
& & 25&1.0020&0.8092&0.7428&0.7195&\textbf{0.7184}\\
& & 50&0.9971&0.8853&0.7865&0.7787&\textbf{0.7766}\\
& & 100&0.9954&1.0426&0.9191&0.9182&\textbf{0.9179}\\
&1&10&0.9996&0.9524&0.8710&\textbf{0.8266}&0.8323\\
& & 25&1.0029&0.9231&0.8937&0.8827&\textbf{0.8811}\\
& & 50&0.9963&0.9868&0.9453&0.9430&\textbf{0.9429}\\
& & 100 &1.0019&1.0024&0.9802&0.9795&\textbf{0.9793}\\
\hline
$\mu_2$&0.5&10&1.0036&0.5693&0.6446&0.5083&\textbf{0.4229}\\
& & 25&1.0010&0.8119&0.7434&0.7165&\textbf{0.6160}\\
& & 50&0.9945&0.9188&0.8338&0.8268&\textbf{0.7535}\\
& & 100&0.9940&0.9752&0.9062&0.9044&\textbf{0.8605}\\
&1&10&0.9983&0.8989&0.8578&0.8006&\textbf{0.7300}\\
& & 25&0.9959&0.9759&0.9155&0.9062&\textbf{0.8612}\\
& & 50&0.9919&0.9927&0.9489&0.9470&\textbf{0.9223}\\
& & 100 &0.9969&0.9949&0.9720&0.9716&\textbf{0.9586}\\
\hline
\end{tabular}
\caption{Empirical risks of the estimators under variance settings with covariance matrix $\Sigma_1$.}
\label{tab1}
\end{table}

\begin{figure}[h]
\includegraphics[width=5.5in]{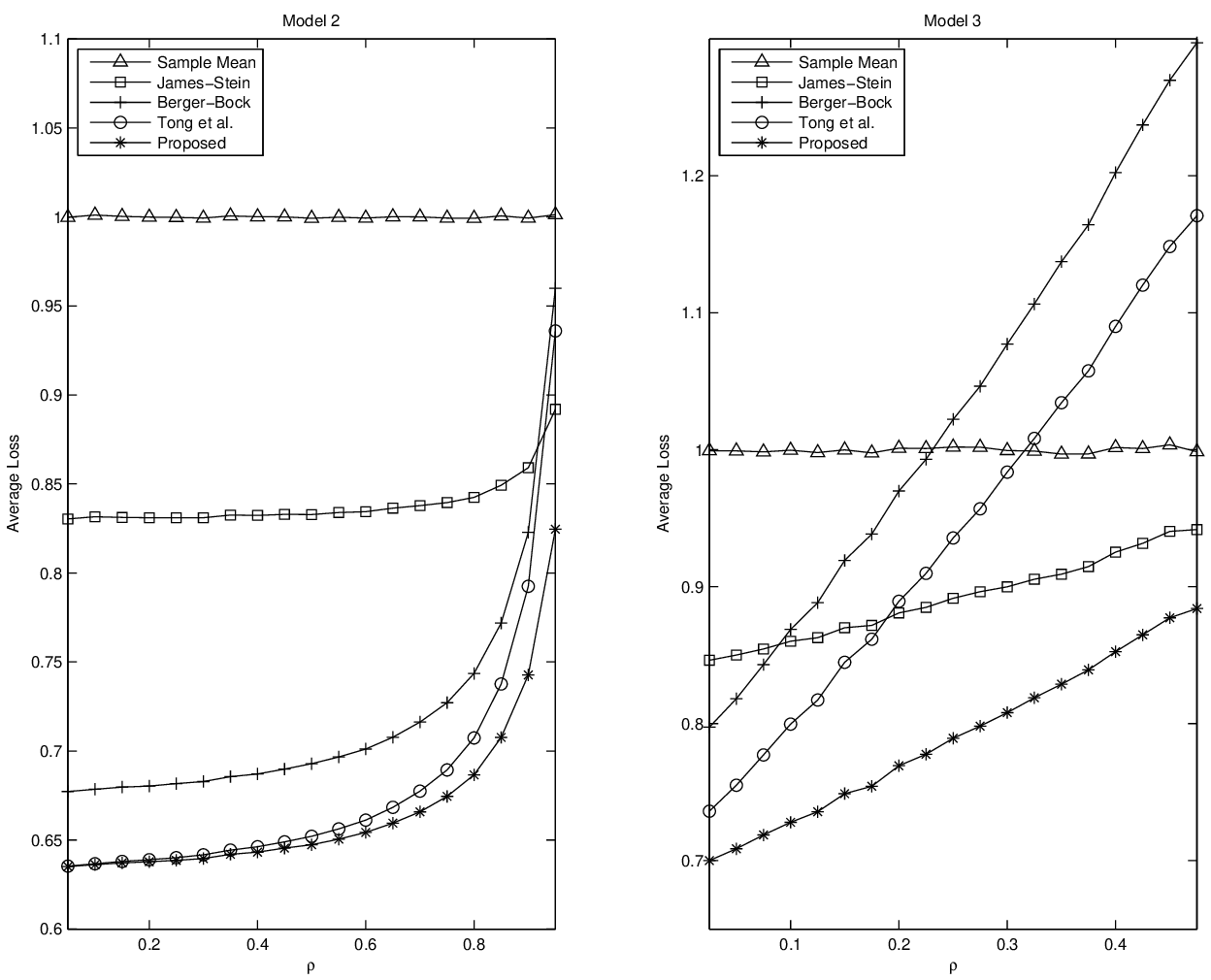}
\caption{Plots of the average losses of the proposed method and existing methods when the observations are correlated. Here $p=100$, $n=20$ and $\mu=\mu_1$ with $\tau=0.5$.}
\label{fig1}
\end{figure}

Finally, we will design simulation studies to evaluate the performance of $\hat{\mu}^*$ with existing methods when the covariance matrix is non-diagonal.
This is to investigate the impact of the correlation coefficient $\rho$ on the performance of the estimators.
To achieve this, we plot in Figure \ref{fig1} the average losses of the estimators for covariance matrices $\Sigma_2$ and $\Sigma_3$ respectively.
To save space, we only present the results for $p=100$, $n=20$, $\mu=\mu_1$ and $\tau=0.5$;
whereas the comparison patterns for other combination settings remain the similar.
From the plots, it is evident that the proposed $\hat{\mu}^*$ provides a smaller average loss than the other estimators in most settings, no matter if $\rho$ is small or not.
We also note that (i) all the shrinkage estimators perform worse when $\rho$ increases;
and (ii) the risks of $\delta_{B}$ and $\delta_{T}$ may be even larger than 1 when the dependence structure is strong, say for $\Sigma_3$ with $\rho\geq 0.35$.

\vskip 12pt
\section{An application}

\noindent
In this section, we illustrate the proposed shrinkage estimator using the Leukemia data in \cite{golub1999molecular}.
The data set contains $p=7129$ genes for 47 acute lymphoblastic leukemia (ALL) and 25 acute myeloid leukemia (AML),
and is available online at the website \emph{http://www.broadinstitute.org/cgi-bin/cancer/datasets.cgi}.

To evaluate the performance of the proposed estimator and compare it with existing methods, we randomly split the 47 ALL samples into the training set $X_1$ and the test set $X_2$.
Specifically, we let the size of the training set range from 5 to 30 and the remaining samples assigned as the test set.
Let $\bar{X}_1$ and $\bar{X}_2$ be the sample means of the training and test sets, respectively.
We further standardize the ALL and AML sets so that each array has variance one across genes.
For simplicity, we let $Q=I_p$ and let the loss function (\ref{risk1}) be $L(\delta|\mu)=(n/p)(\delta-\mu)'(\delta-\mu)$.
Then to compare the performance of the shrinkage estimator $\delta$ and the sample mean $\bar X_1$ based on the training set,
we define the empirical partial risk (EPR) as
\begin{eqnarray}
{\rm EPR} = {L(\bar X_2|\bar X_1) - L(\bar X_2|\delta) \over L(\bar X_2|\bar X_1)} = 1-\frac{|\bar{X}_2-\delta|^2}{|\bar{X}_2-\bar{X}_1|^2}.
\end{eqnarray}
Intuitively, if $\delta$ estimates the true mean $\mu$ more accurately than the sample mean $\bar X_1$,
it will serve as a better proctor of $\mu$ and so $L(\bar X_2|\delta)$ will be smaller than $L(\bar X_2|\bar X_1)$.
As a consequence, if the estimated EPR is larger than 0, we may clarify that $\delta$ is better than $\bar X_1$.
Or equivalently, the EPR may represent the improvement of $\delta$ over $\bar{X}_1$.

\begin{figure}
\centering
\includegraphics[width=5in]{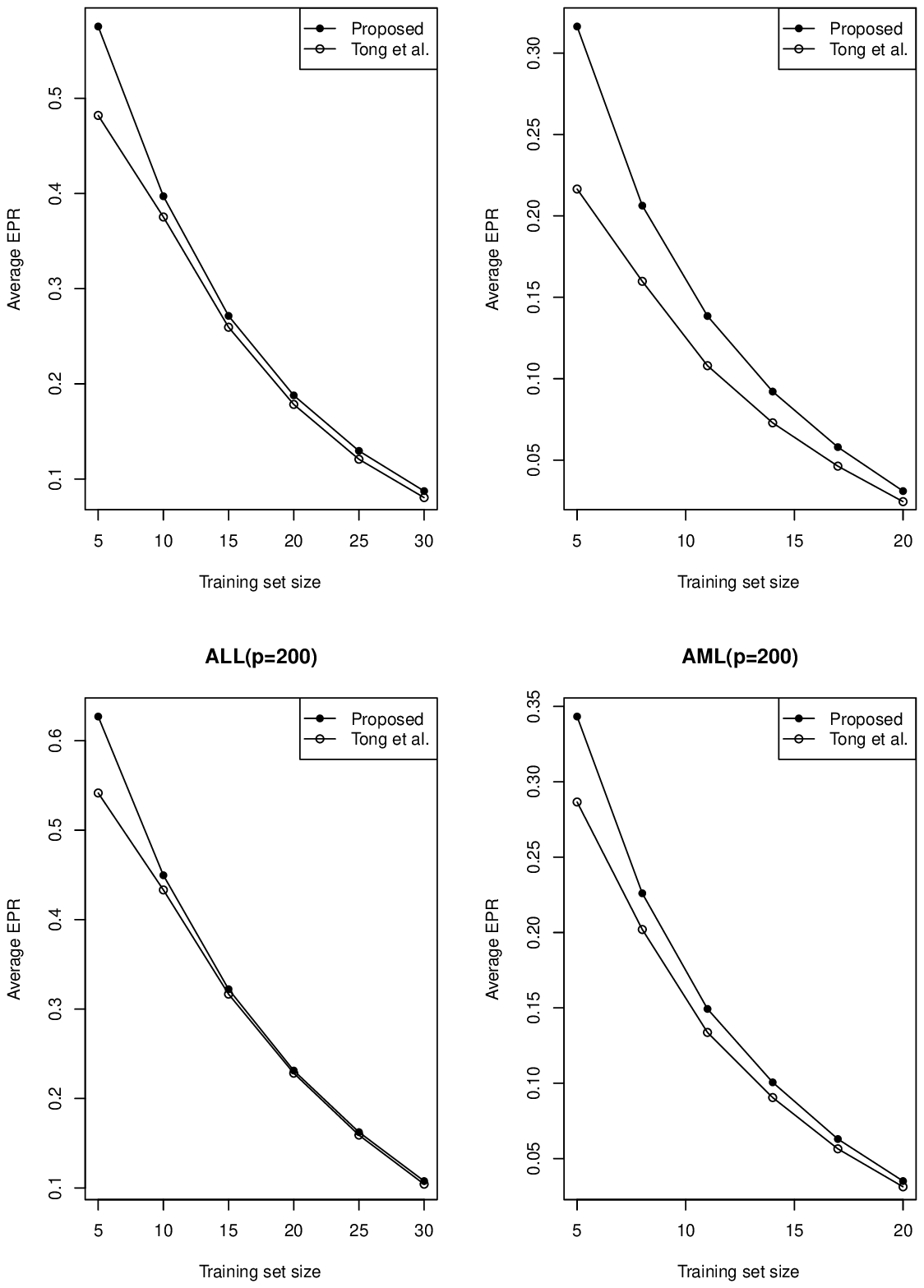}
\caption{The average EPRs of different shrinkage estimators on Leukemia data.}
\label{fig3}
\end{figure}

With $10,000$ simulations, we plot in Figure \ref{fig3} the average EPR using the first 100 and 200 genes of the AML and ALL sets with different sizes of the training set.
Similarly as in Section \ref{part4}, it is evident that
the proposed estimator $\hat{\mu}^*$ outperforms the shrinkage estimator $\delta_T$ in most settings.
We also note that the improvement of $\hat{\mu}^*$ over $\delta_T$ becomes smaller when the size of the training set increases.
This shows that when the sample size is large, the performance of $\hat{\mu}^*$ over $\delta_T$ will be very similar.
Meanwhile, the decreasing pattern of EPR on the training size indicates that both $\hat{\mu}^*$ and $\delta_T$ reduce to the sample mean $\bar X_1$ when the sample size is large.

\vskip 12pt
\section{Conclusion}

\noindent
The paper focuses on the shrinkage mean estimation under the ``large $p$ small $n$" setting.
Specifically, we proposed a shrinkage estimator for the population mean under quadratic loss functions with unknown covariance matrix.
Unlike existing methods in the literature, the proposed method does not assume a specific parametric distribution for the data
and does not require any prior information on the covariance matrix.
In this sense, the proposed estimator is a non-parametric shrinkage estimator and it works for both diagonal and non-diagonal covariance matrices.
Except for the loss function (\ref{loss}), we note that another commonly used quadratic loss function is
\begin{eqnarray} \label{L2}
L(\delta)=(\delta-\mu)^T \Sigma_p^{-1}(\delta-\mu).
\end{eqnarray}
When $\Sigma_p$ is known, it results in a special case of the loss function (\ref{loss}).
For the more realistic setting when $\Sigma_p$ is unknown,
if we do not assume a specific parametric distribution for the data,
a shrinkage estimation for $\mu$ under the loss function (\ref{L2}) would be very difficult to achieve.
Further research is warranted in this direction.

To verify the proposed estimator, we derived some analytical results on the estimator and on the optimal shrinkage coefficients.
The estimators of the optimal shrinkage coefficients were also derived along with some asymptotic properties.
We have also demonstrated through simulation studies using simulated data and real data that the proposed shrinkage estimator performs better
than the sample mean estimation and the existing shrinkage methods under the ``large $p$ small $n$" setting.
Finally, we note that the proposed method (i) extends the methods in \cite{berger1977minimax} and \cite{gleser1986minimax} from the ``small $p$ large $n$" setting
to the ``large $p$ small $n$" setting; and (ii) extends the methods in \cite{berger1976combining} and \cite{tong2012improved}
from a diagonal covariance matrix assumption to a non-diagonal covariance matrix assumption.
The proposed method has extensive applications in different areas including statistical genetics, epidemiology, ecology, and engineering sciences.

\vskip 24pt
\section*{Appendix: Proofs of the Theorems}

\subsection*{A.1. Proof of Theorem \ref{thm1}}

\noindent
By direct calculation, we have
\begin{eqnarray*}
E (\delta-\mu)'Q(\delta-\mu)
&=& \alpha^2 (\mu'Q\mu+\frac{1}{n}{\rm tr}(Q\Sigma_p))+2 \alpha \mu'Q(\beta
e-\mu)+(\beta e-\mu)'Q(\beta e-\mu)\\
&=& \alpha^2 (\mu'Q\mu+\frac{1}{n}{\rm tr}(Q\Sigma_p))- (2 \alpha-1) \mu'Q\mu+\beta^2
e'Q e-2 \beta (1-\alpha) e'Q \mu.
\end{eqnarray*}
This leads to the optimal weights as
\begin{eqnarray*}
&& \alpha^*=\frac{\mu' Q \mu-\frac{(e' Q \mu)^2}{e' Q e}}{\mu'Q \mu +\frac{1}{n}
{\rm tr}(Q \Sigma_p)-\frac{(e' Q \mu)^2}{e'Qe}}\\
&&\beta^* =\frac{e' Q \mu}{e'Q e} (1-\alpha^*).
\end{eqnarray*}
Further, we have
\begin{eqnarray*}
E[L_Q(\mu^*)]&=&\frac{(\mu' Q \mu-\frac{(e' Q \mu)^2}{e' Q
e})}{\mu'Q \mu +\frac{1}{n} {\rm tr}(Q \Sigma_p)-\frac{(e' Q \mu)^2}{e'Qe}}\\
&=& \frac{(\mu-\frac{e' Q \mu}{e' Q e}e)' Q
(\mu-\frac{e' Q \mu}{e' Q e}e)}{(\mu-\frac{e' Q \mu}{e' Q e}e)' Q (\mu-\frac{e'
Q \mu}{e' Q e}e) +\frac{1}{n} {\rm tr}(Q \Sigma_p)}.
\end{eqnarray*}

\vskip 12pt
\subsection*{A.2. Proof of Theorem \ref{lem1}}

\noindent
Without loss of generality, we assume that $E(\epsilon_{11}^4) = 3+\Delta$.  Noting 
\begin{eqnarray*}
X_i=\Sigma_p^{1/2} \epsilon_i+\mu, ~~~~~i=1,\cdots,n,
\end{eqnarray*}
we have 
\begin{align*}
Y_{1,n} &=\frac{n}{p} \mu' Q \mu+\frac{2}{p} \sum_{i=1}^n \epsilon'_i \Sigma_p^{1/2} Q \mu+\frac{1}{p(n-1)} \sum_{i \neq j} \epsilon'_i \Sigma_p^{1/2} Q \Sigma_p^{1/2} \epsilon_j, \\
Y_{2,n} &=\frac{1}{n p} (\sum_{k=1}^n \epsilon'_k \Sigma_p^{1/2} Q \Sigma_p^{1/2} \epsilon_k-\frac{1}{n-1} \sum_{i \neq j} \epsilon'_i \Sigma_p^{1/2} Q \Sigma_p^{1/2} \epsilon_j ),\\
Y_{3,n} &= \frac{n (\mu' Q e)^2}{pe'Qe}+ \frac{1}{ p (n-1)e' Q e } \sum_{i\neq j} e' Q \Sigma_p^{1/2} \epsilon_i \epsilon'_j \Sigma_p^{1/2} Q e+\frac{\mu'Q e}{p e'Q e } \sum_{k=1}^n e'Q \Sigma_p^{1/2} \epsilon_k , \\
Y_{4,n} &= \frac{e'Qu}{e'Qe}+\frac{1}{n e' Q e } \sum_{k=1}^n e' Q \Sigma_p^{1/2} \epsilon_k.
\end{align*}

Then 
\begin{align*}
& E(Y_{1,n}) = \frac{n}{p}\mu' Q \mu,~E(Y_{2,n}) = \frac{1}{p}{\rm tr}(\Sigma_p Q),\\
& E(Y_{3,n}) = \frac{n (\mu' Q e)^2}{pe'Qe},~ E(Y_{4,n}) = \frac{e'Qu}{e'Qe},
\end{align*}
and
\begin{eqnarray*}
&&{\rm Var}(Y_{1,n})=\frac{2 n}{p^2 (n-1)} {\rm tr}(\Sigma_p Q \Sigma_p Q)+\frac{4 n}{p^2} \mu'Q\Sigma_p Q \mu=O(\frac{1}{p})+O(\frac{n \mu'\mu}{p^2}), \\
&&{\rm Var}(Y_{2,n})=\frac{2}{p^2 (n-1)} {\rm tr}(\Sigma_p Q \Sigma_p Q)+\frac{\Delta}{p^2 n} {\rm tr}((\Sigma_p^{1/2}Q\Sigma_p^{1/2})\circ
(\Sigma_p^{1/2}Q\Sigma_p^{1/2}))=O(\frac{1}{n p}),\\
&&{\rm Var}(Y_{3,n})=\frac{2 n}{p^2(n-1)} (\frac{e' Q \Sigma_p Q e}{e' Q e}
)^2+\frac{n (\mu' Q e)^2(e' Q\Sigma Q e)}{p^2 (e'Qe)^2}=O(\frac{1}{p^2})+O(\frac{n \mu'\mu}{p^2}),\\
&&{\rm Var}(Y_{4,n})=\frac{e' Q \Sigma_p Q e}{n (e'Q
e)^2}=O(\frac{1}{ n p}),
\end{eqnarray*}
where $A \circ B=(a_{ij}b_{ij})$ for matrices $A=(a_{ij})$ and $B=(b_{ij})$. This leads to
\begin{eqnarray*}
&&Y_{1,n}=\frac{n}{p} \mu' Q \mu+O_p(\frac{1}{\sqrt{p}}+\sqrt{\frac{n \mu'\mu}{p^2}}),~~~~~Y_{2,n}=\frac{1}{p}
{\rm tr}(\Sigma_p Q)+O_p(\frac{1}{\sqrt{np}}),\\
&&Y_{3,n}= \frac{n (\mu' Q e)^2}{p e' Q e}+O_p(\frac{1}{p}+\sqrt{\frac{n \mu'\mu}{p^2}}),~~~~~Y_{4,n}=\frac{e'Q
u}{e'Q e}+O_p(\frac{1}{\sqrt{np}}).
\end{eqnarray*}
For $\hat{\alpha}^*$ and $\hat{\beta}^*$, we have
\begin{eqnarray*}
\hat{\alpha}^*-\alpha^*&=&\frac{\frac{n}{p}\pi_2+O_p(\frac{1}{\sqrt{p}}+\sqrt{\frac{n \mu'\mu}{p^2}})}{\frac{
n}{p}\pi_1+\frac{n}{p}\pi_2+O_p(\frac{1}{\sqrt{p}}+\sqrt{\frac{n \mu'\mu}{p^2}})}-\frac{\frac{n}{p}\pi_2}{
\frac{n}{p}\pi_1+\frac{n}{p}\pi_2}\\
&=&
\frac{(\frac{n}{p}\pi_1+\frac{n}{p}\pi_2)O_p(\frac{1}{\sqrt{p}}+\sqrt{\frac{n \mu'\mu}{p^2}})}{(\frac{n}{p}\pi_1+\frac{n}{p}
\pi_2+O_p(\frac{1}{\sqrt{p}}+\sqrt{\frac{n \mu'\mu}{p^2}}))(\frac{n}{p}\pi_1+\frac{n}{p}\pi_2)} \\
&=&
\frac{O_p(\frac{1}{\sqrt{p}}+\sqrt{\frac{n \mu'\mu}{p^2}})}{\frac{n}{p}\pi_1+\frac{n}{p}
	\pi_2+O_p(\frac{1}{\sqrt{p}}+\sqrt{\frac{n \mu'\mu}{p^2}})} \\
&=& O_p(\frac{1}{\sqrt{p}}+\sqrt{\frac{n \mu'\mu}{p^2}}),
\end{eqnarray*}
and
\begin{eqnarray*}
\hat{\beta}^* - \beta^*
&=&(1-\hat{\alpha}^*)Y_{4,n} - \beta^*\\
&=& (1-\alpha^*)(Y_{4,n}-\frac{e'Qu}{e'Qe}) -(\hat{\alpha}^*-\alpha^*)Y_{4,n} \\
&=& O_p(\frac{1}{\sqrt{np}}))+(\frac{e'Qu}{e'Qe}+O_p(\frac{1}{
\sqrt{np}})) O_p(\frac{1}{\sqrt{p}}+\sqrt{\frac{n \mu'\mu}{p^2}}) \\
&=& O_p(\frac{1}{\sqrt{np}}+\frac{\sqrt{\mu' \mu}}{p}+\sqrt{n p}(\frac{\sqrt{\mu' \mu}}{p})^2 )=O_p(\frac{1}{\sqrt{np}}+\sqrt{n p}(\frac{\sqrt{\mu' \mu}}{p})^2 ).
\end{eqnarray*}

\vskip 12pt
\subsection*{A.3. Proof of Theorem \ref{thm3}}

\noindent
First consider $(\mu^*-\mu)'Q(\mu^*-\mu)$. Note that
\begin{eqnarray*}
&&(\mu^*-\mu)'Q(\mu^*-\mu)\\
&=& (\alpha^* (\bar{X}-\mu)+\beta^* e+(\alpha^*-1)\mu)'Q (\alpha^*
(\bar{X}-\mu)+\beta^* e+(\alpha^*-1)\mu)\\
&=& (\alpha^*)^2 (\frac{1}{n^2} \sum_{k=1}^n \epsilon'_k \Sigma_p^{1/2}
Q\Sigma_p^{1/2}\epsilon_k+\frac{1}{n^2}\sum_{i \neq j} \epsilon'_i \Sigma_p^{1/2}
Q\Sigma_p^{1/2}\epsilon_j)\\
&&+\frac{2 \alpha^*}{n} (\beta^* e+(\alpha^*-1)\mu)'Q \Sigma_p^{1/2}\sum_{k=1}^n
\epsilon_k+(\beta^* e+(\alpha^*-1)\mu)'Q(\beta^* e+(\alpha^*-1)\mu).
\end{eqnarray*}
Then,
\begin{eqnarray*}
Var((\mu^{*}-\mu)'Q(\mu^{*}-\mu))&\leq & 2 (\alpha^*)^4[\frac{2}{n^2} {\rm tr}(
\Sigma_p Q \Sigma_p Q)+\frac{\Delta}{n^3} \sum_{k=1}^p \phi^2_{kk}] \\
&&+\frac{8(\alpha^*)^2(1-\alpha^*)^2}{n} (\mu-\frac{e'Q \mu}{e'Qe}e)'Q \Sigma_p
Q(\mu-\frac{e'Q \mu}{e'Qe}e),
\end{eqnarray*}
where $\Sigma_p^{1/2}Q\Sigma_p^{1/2}=(\phi_{ij})_{p \times p}$.

By the definitions of $\pi_1$ and $\pi_2$, it is easy to verify that
\begin{eqnarray*}
&&\alpha^*=\frac{\pi_2}{\pi_1+\pi_2},\\
&& E (\mu^{*}-\mu)'Q(\mu^{*}-\mu)=\frac{\pi_1 \pi_2}{\pi_1+\pi_2},\\
&& Var((\mu^{*}-\mu)'Q(\mu^{*}-\mu))\leq \frac{C_1}{n} \frac{\pi_1
\pi_2^4+\pi_1^2 \pi_2^3}{(\pi_1+\pi_2)^4}=(\frac{\pi_1 \pi_2}{\pi_1+\pi_2})^2 O(\frac{1}{p}).
\end{eqnarray*}
Therefore,
\begin{eqnarray}
(\mu^{*}-\mu)'Q(\mu^{*}-\mu)=\frac{\pi_1 \pi_2}{\pi_1+\pi_2}(1+O_p(\frac{1}{\sqrt{p}})). \label{for1}
\end{eqnarray}
By Theorem 3.1, we have
\begin{eqnarray*}
 \hat{\mu}^*=\mu^*+O_p(\frac{1}{\sqrt{p}}+\frac{\sqrt{n \mu'\mu}}{p})
\bar{X}+O_p(\frac{1}{\sqrt{np}}+\frac{\sqrt{n} \mu' \mu}{p \sqrt{p}})e.
 \end{eqnarray*}
Note that
\begin{eqnarray*}
\hat{\beta}^*-\beta^*&=&(1-\hat{\alpha}^*)Y_4-(1-\alpha^*) \frac{e'Qu}{e'Q e}\\
&=& (\alpha^*-\hat{\alpha}^*)\frac{e'Qu}{e'Q
e}+(1-\hat{\alpha}^*)(Y_4-\frac{e'Qu}{e'Qe}),\\
\hat{\beta}^*+\beta^*&=&(1-\hat{\alpha}^*)Y_4+(1-\alpha^*) \frac{e'Qu}{e'Q e}\\
&=& (2-\alpha^*-\hat{\alpha}^*)\frac{e'Qu}{e'Q
e}+(1-\hat{\alpha}^*)(Y_4-\frac{e'Qu}{e'Qe}).
\end{eqnarray*}
We have
\begin{eqnarray}
&&(\hat{\mu}^{*}-\mu)'Q(\hat{\mu}^{*}-\mu)-(\mu^{*}-\mu)'Q(\mu^{*}-\mu)
\nonumber \\
&=& (\hat{\mu}^*-\mu^*)'Q(\hat{\mu}^*+\mu^*-2 u)\nonumber \\
&=& ((\hat{\alpha}^*-\alpha^*)\bar{X}+(\hat{\beta}^*-\beta^*)e)'Q
((\hat{\alpha}^*+\alpha^*)\bar{X}+(\hat{\beta}^*+\beta^*)e-2 u)\nonumber \\
&=&
((\hat{\alpha}^*-\alpha^*)(\bar{X}-\mu)+(\alpha^*-\hat{\alpha}^*)(\frac{e'Qu}{
e'Q e}e-\mu)+(1-\hat{\alpha}^*)(Y_4-\frac{e'Qu}{e'Qe})e)'Q \nonumber \\
&&
*((\hat{\alpha}^*+\alpha^*)(\bar{X}-\mu)+(2-\hat{\alpha}^*-\alpha^*)(\frac{e'Q
\mu}{e'Qe}e-\mu)+(1-\hat{\alpha}^*)(Y_4-\frac{e'Qu}{e'Qe})e)\nonumber \\
&=&
(\hat{\alpha}^*-\alpha^*)[(\hat{\alpha}^*+\alpha^*)(\bar{X}-\mu)'Q(\bar{X}
-\mu)+2 (1-\hat{\alpha}^*-\alpha^*)(\bar{X}-\mu)'Q (\frac{e'Q
\mu}{e'Qe}e-\mu)]\nonumber \\
&& + 2\hat{\alpha}^*(1-\hat{\alpha}^*)(Y_4-\frac{e'Qu}{e'Qe})(\bar{X}-\mu)'Q e
+(Y_4-\frac{e'Qu}{e'Qe})^2(1-\hat{\alpha}^*)^2 e'Qe \nonumber \\
&&+ (\hat{\alpha}^*-\alpha^*) (2-\hat{\alpha}^*-\alpha^*) (\mu-\frac{e'Qu}{e'Q
e}e)'Q  (\mu-\frac{e'Qu}{e'Q e}e) \nonumber \\
&=&\pi_1
O_p(\frac{1}{\sqrt{p}}+\frac{\sqrt{n \mu'\mu}}{p})+O_p(\frac{1}{\sqrt{p}}+\frac{\sqrt{n \mu' \mu}}{p})
O_p(\sqrt{\frac{\mu' \mu}{n}}) \nonumber \\
&&+O_p(\frac{1}{n})+\pi_2 O_p(\frac{1}{\sqrt{p}}+\frac{\sqrt{n \mu'\mu}}{p})(\frac{2 \pi_1}{\pi_1+\pi_2}+\frac{O_p(\frac{1}{\sqrt{p}}+\frac{\sqrt{n \mu'\mu}}{p})}{\frac{n}{p}\pi_1+\frac{n}{p}\pi_2+O_p(\frac{1}{\sqrt{p}}+\frac{\sqrt{n \mu'\mu}}{p})}) \nonumber \\
&=& \pi_1
O_p(\frac{1}{\sqrt{p}}+\frac{\sqrt{n \mu'\mu}}{p}),
\label{for2}
\end{eqnarray}
where we used the facts that
\begin{eqnarray*}
&&(\bar{X}-\mu)'Q(\bar{X}-\mu)=\pi_1(1+O_p(\frac{1}{\sqrt{p}})),\\
&&(\bar{X}-\mu)'Q (\frac{e'Q \mu}{e'Qe}e-\mu)=O_p(\sqrt{\frac{\mu'\mu}{n}}),\\
&&(\bar{X}-\mu)'Q e=O_p(\sqrt{\frac{p}{n}}).
\end{eqnarray*}
Finally, by (\ref{for1}) and (\ref{for2}) we have
\begin{eqnarray}
(\hat{\mu}^{*}-\mu)'Q(\hat{\mu}^{*}-\mu)=\pi_1 (\frac{
\pi_2}{\pi_1+\pi_2}+O_p(\frac{1}{\sqrt{p}}+\frac{\sqrt{n \mu' \mu}}{p})).
\end{eqnarray}
This completes the proof of Theorem \ref{thm3}.

\vskip 12pt
\noindent
{\it Conflict of Interest}: None declared.

\bibliographystyle{natbib}
\bibliography{cit}

\end{document}